\documentclass[12pt]{article}
\usepackage{amsfonts}
\usepackage{amsmath}
\usepackage{graphicx}
\usepackage{amssymb}

\numberwithin{equation}{section}

\newtheorem{theorem}{Theorem}[section]
\newtheorem{lemma}{Lemma}[section]

\newtheorem{definition}{Definition}

\newtheorem{remark}{Remark}[section]

\def\rd{\rm d}

\def\td{\tt d}
\def\tG{\tt G}
\def\tg{\tt g}
\def\tT{\tt T}
\def\ttg{\tt g}
\def\ttG{\tt G}
\def\bttg{\overline{\tt g}}

\def\boeta{{\mbox{\boldmath$\eta$}}}
\def\oboeta{\overline\boeta}

\def\om{\omega}
\def\oom{\overline\omega}
\def\bttg{\mbox{\boldmath${\tt g}$}}

\def\bom{{\mbox{\boldmath$\omega$}}}
\def\obom{\overline\bom}
\def\0bom{{\bom}^0}
\def\0obom{{\obom}^0}

\def\0nbom{{\bom}_{n,0}}
\def\n*bom{{\bom}^*_{(n)}}

\def\oom{\overline\om}

\def\bOm{\mbox{\boldmath${\Omega}$}}

\def\utheta{\underline\theta}
\def\ovr{\overline r}

\def\bbC{\mathbb C}

\def\bbP{\mathbb P}
\def\fB{\mathfrak B}

\def\bx{\mathbf x}

\def\cE{\mathcal E}

\def\cH{\mathcal H}

\def\cT{\mathcal T}
\def\cV{\mathcal V}

\def\ocH{\overline\cH}

\def\bbR{\mathbb R}

\def\bbZ{\mathbb Z}

\def\bx{\mathbf x}

\def\by{\mathbf y}

\def\unn{\underline n}

\def\ovp{\overline p}
\def\bx{\mathbf x}
\def\ox{\overline x}

\def\diy{\displaystyle}
\def\ov{\overline}

\def\oV{\overline {\mathcal V}}

\def\oW{\overline W}

\def\LT{{\mathbb{LT}}}

\def\rd{\rm d}

\begin{document}
\title{A Mermin--Wagner theorem\\ on Lorentzian triangulations\\
with quantum spins}
\author{ M. Kelbert$^{1,3}$ \and Yu. Suhov~$^{2,3}$\and A. Yambartsev~$^{3}$}
\vspace{1mm}

\maketitle {\footnotesize
\noindent $^{1}$ Department of Mathematics, Swansea University, UK\\
E-mail: M.Kelbert@swansea.ac.uk

\noindent $^2$ Statistical Laboratory, DPMMS, University of Cambridge, UK\\
E-mail: I.M.Soukhov@statslab.cam.ac.uk

\noindent $^3$ Department of Statistics, Institute of Mathematics
and Statistics, \\ University of S\~ao Paulo, Brazil.\\
E-mail: yambar@ime.usp.br }

\begin{abstract}
We consider infinite random causal Lorentzian triangulations emerging
in quantum gravity for critical values of parameters. With each
vertex of the triangulation we associate a Hilbert space representing a
bosonic particle moving in accordance with the standard laws of Quantum Mechanics.
The particles interact via two-body potentials decaying with the graph distance.
A Mermin--Wagner type theorem is proven for infinite-volume reduced
density matrices
related to solutions to DLR equations in the Feynman--Kac (FK) representation.
\vskip.1truecm

\end{abstract}
\newpage

\section{Introduction}

In this paper we prove a Mermin-Wagner (MW) type theorem (cf. Mermin and Wagner (1966), Dobrushin and Shlosman (1975), Ioffe et al. (2002)) for a system
of quantum bosonic particles on an (infinite) random graph represented by a {\it causal dynamical
Lorentzian triangulation} (in brief: CDLT). The CDLTs arise naturally
when physicists attempt to define a fundamental path integral in quantum gravity.
The reader is referred to Loll et al. (2006) for a review of related publications and to
Malyshev et al. (2001) for a rigorous mathematical background behind the model of CDLTs.
More precisely, we analyze a quantum system on a random 2D graph $T$ generated
by a natural ``uniform"  measure on the CDLTs corresponding to a ``critical" regime
(see below).

In modern language, the spirit of the quantum MW theorem is that in a
 2D lattice model (more generally, for a model on a countable bi-dimensional graph),
any infinite-volume Gibbs state (regardless of whether it is unique or not)
is invariant under the action of a Lie group ${\tG}$ provided that the
ingredients of local Hamiltonians are ${\tG}$-invariant; see Mermin and Wagner (1966).
These ingredients include the kinetic energy part, the single-site potential
and the interaction potential. The mathematical
constructions used for the proof of this theorem
require a certain control over these ingredients: compactness of
a configuration space associated with a single vertex of the lattice
(or a more general bi-dimensional graph), a certain smoothness of the interaction
potential (or its essential part), sufficiently fast decay of the interaction potential
for large distances on the lattice (or on the graph), ``regularity" of
the lattice (graph) geometry. In particular, the bi-dimensionality of
the underlying graph is guaranteed by Eqn (4.1.1).

A principal question that needs a careful consideration is
about the definition of a quantum Gibbs state in an infinite volume.
For the so-called quantum spin systems, with a finite-dimensional phase
space of a single spin (and consequently, with bounded local
Hamiltonians), such a definition is given within the theory
of KMS (Kubo--Martin--Schwinger) states; see, e.g., Bratteli and Robinson (2002). A version
of the MW theorem for a model of this type on a 2D square lattice was
established in Fr\"ohlich and Pfister (1981), Pfister (1981) and has been generalised  in 
subsequent publications. The
KMS-based results (under suitable aforementioned assumptions)
can be extended to the model of
classical spins on a random graph of the type considered in the
present paper; cf. Kelbert et al. (2013, \cite{KSY}). However, the KMS-theory is not efficient
for the case of interacting quantum particles where the one-particle
kinetic energy operator is equal to $-\Delta /2$ ($\Delta$ stands for
a Laplacian on a compact manifold). This is a standard quantum-mechanical
model, and the fact that the concept of infinite-volume Gibbs state did not
receive so far a properly working definition for such a system was
perceived as a regrettable hindrance.

In this paper we adopt the definition of an infinite-volume Gibbs state
(more precisely, of an infinite-volume reduced density matrix (RDM, for short))
from the papers Kelbert and Suhov, 2013 (cf. \cite{KS1}, \cite{KS2}). Similar methodologies
have been developed in a number of earlier references; see, e.g., Albeverio et al. \cite{AKKR} and
the bibliography therein (viz., Klein and Landau \cite{KL}). In papers Kelbert and Suhov \cite{KS1}, \cite{KS2} 
a class of so-called FK-DLR states
has been introduced, and an MW theorem was established for quantum systems
on a bi-dimensional graph $T\sim (\cV,\cE)$ where $\cV =\cV(T)$ is the
set of vertices and $\cE=\cE (T)$ the set of edges. As
was said above, in the present paper we deal with
a random graph $T$ (a CDLT for critical values of parameters). After
checking that a typical realization $T=T_{\infty}$ of the random CDLT
satifies certain required properties, we use the constructions from Kelbert and Suhov \cite{KS1},
\cite{KS2}) (going back to Fr\"ohlich and Pfister (1981), Pfister (1981)) and prove the main results
of the paper (see Theorems \ref{main1}, \ref{main2} and Theorems
\ref{Main1}, \ref{Main2}).

It is appropriate to say that, although we use here some methodology
developed in Kelbert and Suhov (2013, \cite{KS1} and \cite{KS2}), the current work deals with a
situation  different from the above papers, and a number of issues here
require specific technical tools. On the other hand, the present paper
can be considered  as a continuation of Kelbert et al. (2013, \cite{KSY}) where a MW theorem
was established for a classical prototype of a quantum system treated here.
We believe that models of quantum gravity where various types of quantum
matter are incorporated form a natural direction of research, interesting
from both physical and mathematical points of view. Extension of results from
Kelbert et al. (2013, \cite{KSY}) to the case of quantum systems is a novel element of
the present paper.

\newpage

\section{Basic definitions}
\vskip 1 truecm

{\bf 2.1. Lorentzian triangulations in a critical phase.}

\begin{figure}[!ht]
  \centering
   \scalebox{0.7}{\includegraphics{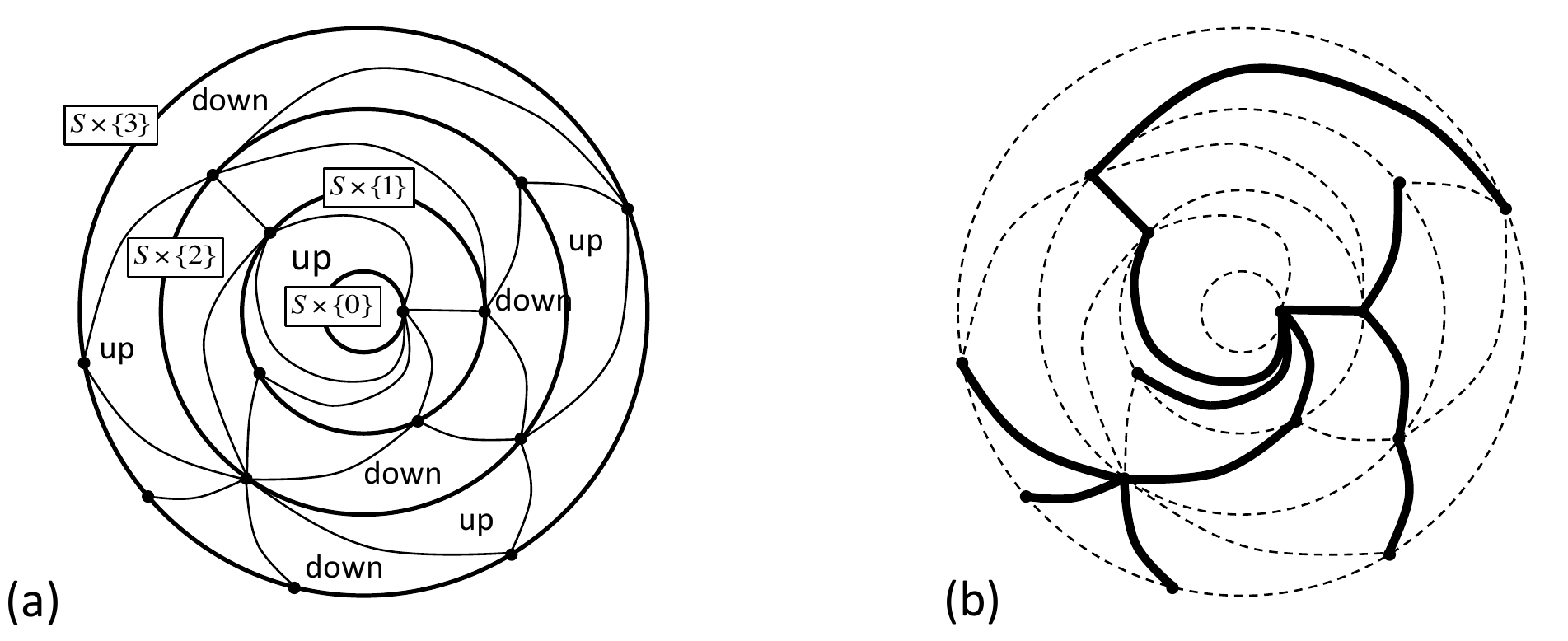}}
  \caption{(a) An example of Lorentzian triangulation. Some (not all) up- and down-triangles are marked.
  (b) The triangulation $T$ is parametrized by the spanning tree $\tT$, which is represented by bold lines.}
  \label{fig1}
\end{figure}

The graph under
consideration is a triangulation $T$ of a cylinder
${\sf C}=S\times [0,\infty)$ with the base $S$, which is a unit
circle in $\bbR^2$. Physically, ${\sf C}$ represents a $(1+1)$-dimensional
space-time complex. (Pictorially, in the critical case, the graph $T$ develops like a cone, 
getting ``wider" further from the base.)  Geometrically, ${\sf C}$ can be visualized as
a complex plane $\bbC$ with a family of concentric circles $\{z:\;|z|=n\}$:
here the origin $z=0$ is treated as a ``circle" of infinitesimal radius.
The following properties of $T$ are assumed: each triangle belongs to
some strip
$S\times [\ell ,\ell+1]$, $\ell =0,1,2,\ldots$ such that either (i)
two vertices lie on $S\times\{\ell\}$
and one on $S\times\{\ell+1\}$ or (ii) two vertices lie on $S\times\{\ell +1\}$
and one on $S\times\{\ell\}$ and exactly one edge of triangle is an arc of a circle $S\times\{\ell\}$ in the case (i), or
$S\times\{\ell+1\}$ in the case (ii). In case (i) we speak of an upward triangle, or simply up-triangle, and in case (ii)
of an downward triangle, or down-triangle. This includes also a ``degenerate" picture where two vertices of a
triangle coincide, and the corresponding edge forms the circle.
For $\ell =0$ it is requested that the graph under consideration
generates a degenerate picture (i.e., the graph has a single up-triangle in the strip
$S\times [0,1]$, see Figure~\ref{fig1}(a). This particular
triangle is called the root triangle, and its side represented by the edge
along the boundary $S\times\{0\}$ of the strip is called the root edge. Moreover, the (double)
vertex lying on $S\times\{0\}$ is called the root vertex.

Finally, we consider graphs modulo an equivalence (that is, up to a homeomorphism
of ${\sf C}$ preserving all circles $S\times\{\ell\}$, $\ell =0,1,2,\ldots$).

\begin{definition} A rooted infinite CDLT is defined as an equivalence class
of (countable) graphs with the above-listed properties, under the above equivalence.
Depending on the context, we use the notation $T=T_\infty$ for
a representative or for the whole
equivalence class of graphs involved, and speak of the vertex set
$\cV=\cV(T)$ and the edge set $\cE=\cE(T)$ in the same fashion.
\end{definition}
\medskip

A similar definition can be used to introduce a rooted CDLT on a
cylinder ${\sf C}_N=S\times [0,N]$ (a rooted CDLT
of height $N$). The corresponding notation is $T_N\sim (\cV(T_N),
\cE(T_N))$ or even $T_N\sim (\cV_N,\cE_N)$. As in Kelbert et al. (2013, \cite{KSY}),
we denote by $\LT_N$ and
$\LT_{\infty}$ the sets of CDLTs on ${\sf C}_N$ and ${\sf C}$
respectively.

To introduce a probability distribution on $\LT_N$ and ultimately on
$\LT_{\infty}$, we use a special 1-1 correspondence between
the rooted CDLTs and rooted trees (that is, graphs without cycles
and with distinguished vertices). Namely, we extract
a subgraph in $T$ by selecting, for each vertex $v\in T$, the leftmost edge
going from $v$ downwards
and discarding all other edges going from $v$ horizontally or downwards, see Figure~\ref{fig1}(b).
The graph ${\tT}\subset T$ thus obtained is a spanning tree of $T$, cf. Durhuus et al. (2006) and Malyshev et al. (2001).
Moreover, if one indicates, for each vertex of ${\tT}$, its height in $T$
then $T$ can be completely reconstructed when we know $\tT$.
We call the correspondence ${\tT}\leftrightarrow T$ the {\it tree parametrization}
of the CDLT.
It determines a one-to-one bijection ${\sf m}$ between the set
$\LT_\infty$ and the set of infinite rooted trees $\mathcal T_\infty$:
$$
{\sf m}: \mathcal T_\infty \leftrightarrow  \LT_\infty .$$
We will use the same symbol ${\sf m}$ for the bijection $\mathcal T_N \leftrightarrow  \LT_N$
where $\mathcal T_N$ is the set of all rooted planar trees of height $N$.

By virtue of the tree-parametrization, we will specify a probability distribution
on CDLTs by specifying a distribution defined on trees. More precisely, suppose
${\sf P}^{\rm{tree}}_N$ is a probability measure
on $\mathcal T_N$. Then the measure ${\sf P}^{\rm{LT}}_N$ on $\LT_N$ is determined by
$$
{\sf P}^{\rm{LT}}_N ( {\sf m}({\tT}) ) =  {\sf P}^{\rm{tree}}_N ({\tT}),\;\;
\forall\;\;{\tT}\in \mathcal T_N.
$$
Conversely, let ${\sf P}^{\rm{tree}}$ be a probability distribution on
$\mathcal T_\infty$. Then the distribution ${\sf P}^{\rm{LT} }$ on $\LT_\infty$
is given by
$${\sf P}^{\rm{LT}}( {\sf m}(A) )=  {\sf P}^{\rm{tree}}( A ),\;\;
\forall\;\;A \in {\mathcal F}(\mathcal T_\infty),$$
where ${\mathcal F}(\mathcal T_\infty)$ is standard $\sigma$-algebra generated by
cylinder sets.
In future we omit indices in the notation for the distribution $\sf P$.

To construct a critical CDLT model we define
the corresponding measure on $\mathcal T_\infty$ related to a critical Galton-Watson (GW)
process $\xi$. For this aim, we set $\mu = \{p_k\}$
to be an offspring distribution on
$\mathbb N = \{0,1,\dots \}$ with mean 1, and define the critical
Galton--Watson (GW) branching process $\xi$. Conditional on the
event of non-extinction, the GW process becomes
the so-called size-biased (SB) process ${\hat\xi}= \{k_n$, $n=0,1,2,\ldots\}$. In
our context, $k_n$ yields the number of vertices on the circle
$S\times \{n\}$ in the random infinite rooted CDLT. The reader can
consult Lyons et al. (1995) for the formal background for SB branching processes.

In particular, the distribution of an SB process $\hat\xi$ is concentrated on the subset
$\mathcal S$ of ${\mathcal T}_\infty$ formed by the so-called single-spine trees. A
single-spine tree consists of a single infinite linear chain $s_0, s_1, \dots $
called the spine, to each vertex $s_j$ of which there is attached a finite
random tree with its root at $s_j$. (Here $s_0$ is the root vertex
of the whole tree.)
Furthermore, the
generating function for the branching number $\nu$ at each vertex $s_j$
is $f'(x)$ where $f(x)$ is the generating function of the initial
offspring distribution $\mu$.
Moreover, the individual branches are independently and identically
distributed in accordance with the original critical GW process Lyons et al. (1995).

Let $\sigma^2$ stand for the variance of the offspring distribution $\nu$. Then
$$
{\sf E} (k_n \mid k_{n-1}) = k_{n-1} + \sigma^2.
\eqno (2.1.1)$$
In fact, let $\nu = \{\tilde p_k\}$ be the SB offspring distribution with
$\tilde p_k = k p_k$ (recall, in the critical case under consideration,
the sum $\sum\limits_kkp_k=1$). Then the distribution for $k_n$ conditioned upon the
value $k_{n-1}$ is identified as follows (cf. Lyons et al. (1995)). We choose at random
one particle among $k_{n-1}$ particles and generate the number of its
descendants according to the distribution $\nu$ with mean $\sigma^2 +1$.
According to (2.1.1), the number of descendants for the other $k_{n-1}-1$ particles is generated,
independently, by the distribution $\mu$.

Throughout the paper we assume that the offspring distribution $\mu$ has the
mean 1 with finite second moment. Let $\sf P$
be the corresponding SB Galton-Watson tree distribution.
\medskip\medskip

{\bf 2.2. The local quantum Hamiltonians on CDLTs.} Let $M=\bbR^d/\bbZ^d$
be a unit $d$-dimensional
torus with flat metric and induced volume $v$. A basic quantum
model uses the Hilbert space $\cH ={\rm L}_2(M,v)$ as the phase space of a single
quantum particle. The single-particle Hamiltonian $H$ acts in $\cH$ as the
sum:
$$(H\phi)(x)=-
\frac{1}{2}(\Delta \phi )(x)+U(x)\phi (x),\;\;x\in M,\;\phi\in\cH.\eqno{(2.2.1)}$$
Here $\Delta$ is the Laplacian on $M$ and the function $U:x\in M\mapsto \bbR$ gives an
external potential. Under the assumptions upon $U$ adopted in this paper (see Eqn (2.3.1)),
$H$ is a self-adjoint operator bounded from below and with a discrete spectrum such
that $\forall$ $\beta >0$, $\exp\,[-\beta H]$ is a (positive definite) trace class
operator.

Given a CDLT $T$ we use the notation $T_N$ for the subgraph in $T$
with the set of vertices $\cV_N=\cV(T_N)$ of the form
$\cV_N=\cV(T)\cap\{S\times\{0,\ldots N\}\}$ and the set of edges
$\cE(T_N)=\cE (T)\cap (\cV_N\times\cV_N)$. The phase space of a (bosonic)
quantum system in $\cV_N$ is the tensor product $\cH_N=\cH^{\otimes \cV_N}$;
an element $\phi_N\in\cH_N$ is a function
$$\phi_N:\,\bx(N)=\bx_{\cV_N}=\{x(i),i\in \cV_N\}\in M^{\times\cV_N}\mapsto\bbC\eqno (2.2.2)$$
square-integrable in ${\rm d}\bx (N)=\prod\limits_{i\in\cV_N}v({\rm d}x(i))$.
The Cartesian power $M^{\times\cV_N}$ can be considered as the
configuration space for the classical prototype of the quantum system in $\cV_N$.

The local Hamiltonian
$H_N$ of the system in $\cV_N$ acts on functions $\phi\in
\cH^{\otimes\cV_N}$: given $\bx(N)=(x(j),\,j\in \cV_N )
\in M^{\times\cV_N}$,
$$\big(H_N\phi\big)(\bx(N))=
\left[\sum_{i\in \cV_N}H(i)+
\sum_{j,j'\in \cV_N\times \cV_N} J({\td}(j,j'))\,V(x(j),x(j'))\right]
\phi (\bx(N)),\eqno{(2.2.3)}$$
where $J({\td}(j,j'))\,V(x(j),x(j'))$ represents the interaction between
spins  $x(j)$ and $x(j')$ at sites $j$ and $j'$.
Next, $H(i)$ stands for the copy of operator $H$ acting on
variable $x(i)\in M$ and ${\td}(j,j')$ for the graph distance
from vertex $j$ to $j'$.

A more general concept is a Hamiltonian
$H_{N|\,\bx^{\rm c} (N)}$ in the external field generated by an
(infinite) configuration $\bx^{\rm c}(N)
=\{\ox (j'),\,j'\in\oV_N\}\in M^{\times\oV_N}$ where $\oV_N=\cV(T)\setminus \cV_N$.
As before, operator $H_{N |\,\bx^{\rm c}(N)}$
acts in $\cH_N$: given $\phi\in\cH_N$ and
$\bx (N)=(x(j),\,j\in \cV_N )\in M^{\times\cV_N}$,
$$(H_{N |\,\bx^{\rm c}(N)}\phi\big)(\bx(N))=
\bigg[H_N+\sum\limits_{(j,j')\in\cV_N\times\oV_N}
J({\td}(j,j'))\,V(x(j),\ox(j'))\bigg]
\phi (\bx(N)).\eqno{(2.2.4)}$$
Again, under assumptions upon $J$ and $V$ described in (2.3.1)-(2.3.2),
$H_N$ and $H_{N |\,\bx^{\rm c}(N)}$ are self-adjoint operators bounded
from below and with a discrete spectrum such
that $\forall$ $\beta >0$, $G_{\beta ,N}=\exp\,[-\beta H_N]$ and
$G_{\beta ,N |\,\bx^{\rm c}(N)}=\exp\,\big[-\beta H_{N |\,\bx^{\rm c}(N)}\big]$
are (positive definite) trace class operators.

The operators $G_{\beta ,N}$ and $G_{\beta ,N |\,\bx^{\rm c}(N)}$ are called
the Gibbs operators (in volume $\cV_N$ for the inverse temperature $\beta$
and, in the case of $G_{\beta ,N |\,\bx^{\rm c}(N)}$, with the boundary condition
$\bx^{\rm c}(N)=\bx^{\rm c}_{\oV_N}$).
The traces
$$\Xi_{\beta ,N}={\rm{tr}}_{\cH_N}G_{\beta ,N}
\;\hbox{ and }\;\Xi_{\beta ,N|\,\bx^{\rm c}(N)}={\rm{tr}}_{\cH_N}G_{\beta ,N |\,\bx^{\rm c}(N)}
\eqno{(2.2.5)}$$
give the corresponding partition functions. The normalized operators
$$R_{\beta ,N} =\frac{1}{\Xi_{\beta ,N}}\,G_{\beta ,N}
\;\hbox{ and }\;R_{\beta ,N |\,\bx^{\rm c}(N)}=\frac{1}{\Xi_{\beta ,N|\,\bx^{\rm c}(N)}}
\,G_{\beta ,N |\,\bx^{\rm c}(N)}\eqno{(2.2.6)}$$
are called the density matrices (for the corresponding Gibbs ensembles); these
are positive definite operators of trace $1$. Given $n\in\{0,\ldots ,N\}$,
the partial traces
$$R^{(n)}_{\beta ,N}={\rm{tr}}_{\,\cH_{N\setminus n}}R_{\beta ,N}\;\hbox{ and }
R^{(n)}_{\beta ,N |\,\bx^{\rm c}(N)}={\rm{tr}}_{\,\cH_{N\setminus n}}R_{\beta ,N |\,\bx^{\rm c}(N)}
\eqno{(2.2.7)}$$
yield positive definite operators $R^{(n)}_{\beta ,N}$ and
$R^{(n)}_{\beta ,N |\,\bx^{\rm c}(N)}$ in $\cH_{n}$, of trace $1$. Here
$\cH_{N\setminus n}$ stands for the Hilbert space $\cH^{\otimes (\cV_N\setminus\cV_{n})}$.
These operators are called the reduced density matrices (RDMs). Note the compatibility
relation: $\forall$ $0\leq n<n'<N$:
$$R^{(n)}_{\beta ,N}={\rm{tr}}_{\,\cH_{n'\setminus n}}R^{(n')}_{\beta ,N}\;\hbox{ and }
R^{(n)}_{\beta ,N |\,\bx^{\rm c}(N)}={\rm{tr}}_{\,\cH_{n'\setminus n}}R^{(n')}_{\beta ,N |\,\bx^{\rm c}(N)}
\eqno{(2.2.8)}$$

\vskip .5 truecm

{\bf 2.3. Assumptions on the potentials. The group of symmetries.}
We suppose that the potential $U$ has continuous derivatives
whereas $V$ has continuous first and second derivatives:
$\forall$ $x,x',x''\in M$
$$|U(x)|, \left|\nabla_x U(x)\right|\leq {\ov U},\eqno{(2.3.1)}$$
$$
|V(x',x'')|, \left|\nabla_{{x}'}V(x',x'')\right|, \left|\nabla_{{x}''}V(x',x'')\right|,
\left|\nabla_{{x}'}\nabla_{{x}''}V(x',x'')\right|
\leq {\ov V},\eqno{(2.3.2)}$$
where ${\ov U},{\ov V}\in (0,\infty )$ are constants.

Next, suppose that a $d'\times d$ matrix $A$ is given, of
the row rank $d'$ where $d'\leq d$. We consider a $d'$-dimensional
group ${\tG}$ acting on $M$ and preserving the volume $v$:
$({\tg},x)\in{\tG}\times M\mapsto {\tg}x\in M$. More precisely,
${\tg}$ is identified with a real $d'$-dimensional vector
$\theta =(\theta_1,\ldots ,\theta_{d'})$
and the action is given by
$${\tg}x=x+\theta A\;{\rm{mod}}\;1.\eqno{(2.3.3)}$$
\vskip .1 truecm

\begin{remark} The group
${\tG}$ can be compact (in which case ${\tG}$ is a torus of dimension $d'$)
or non-compact (then ${\tG}$ is $\bbR^{d'}$).
\end{remark}
\vskip .1 truecm

We assume that the functions $U(x)$ and $V(x,x')$ are invariant
with respect to the group $\tG$:
$\forall {\tg}\in {\tG}$ and $x,x'\in M$
$$U({\tg}x)=U(x),\;\; V({\tg}x,{\tg}x')=V(x,x').\eqno{(2.3.4)}$$

Finally, we assume that the function $r\in (0,\infty )\mapsto J(r)$
in (2.2.3) and (2.2.4) is a bounded monotone decreasing function satisfying the
condition
$$
J(r) \le \Bigl( \frac{1}{r \ln r} \Bigr)^3,\;\; r\geq 2.\eqno{(2.3.5)}
$$

These assumptions are in place throughout the paper. (We do not analyze the issue
of necessity of condition (2.3.5).)

As usually, the action of the group ${\tG}$ generates unitary operators in
$\cH$:
$$S({\tg})\phi (x)=\phi ({\tg}^{-1}x),\;\;x\in M,\;\phi\in\cH.
\eqno (2.3.6)$$
Let $S^{(N)}({\tg})$ be the tensor power of $S({\tg})$ which acts in $\cH_N$:
for any $\phi_N\in\cH_N$
$$S^{(N)}({\tg})\phi_N(\bx (N))=\phi_N({\tg}^{-1}\bx (N)),\eqno{(2.3.7)}$$
where $
\bx (N)=\{x(i),\;i\in\cV_N\}\in M^{\times\cV_N}
$
and ${\tg}^{-1}\bx (N)=\{{\tg}^{-1}x(i),\;i\in\cV_N\}$.
\vskip .5 truecm

{\bf 2.4. Limiting RDMs in an infinite volume.} We are interested in
the `thermodynamic' limit $N\to\infty$. In the absense of phase transitions, 
one would like to
establish a convergence of the RDMs $R^{(n)}_{\beta ,N}$ and
$R^{(n)}_{\beta ,N |\,\bx^{\rm c}(N)}$ to a limiting RDM in $\cH_{n}$ as $N\to\infty$.
A suitable form of convergence is in the trace norm
in $\cH_{n}$, guaranteeing that the limiting operator is positive-definite
and has trace $1$. When phase transitions are not excluded 
(which is the case under consideration), a more general question is whether the families
$\{R^{(n)}_{\beta ,N}\}$ and $\{R^{(n)}_{\beta ,N |\,\bx^{\rm c}(N)}\}$
are compact.  If we manage to check that $\{R^{(n)}_{\beta ,N}\}$
and $\{R^{(n)}_{\beta ,N |\,\bx^{\rm c}(N)}\}$ are compact families for
any given $n$ then, invoking a diagonal process, we can consider
a family of limiting RDMs $\{R^{(n)}_\beta,\;n=0,1,2,\ldots\}$ (in the case
of operators $R^{(n)}_{\beta ,N |\,\bx^{\rm c}(N)}$ the limiting RDMs may depend
on the choice of the boundary conditions $\bx^{\rm c}(N)$).
The consistency property (2.2.8) will be inherited in the limit:
$\forall$ $0\leq n<n'<N$,
$$R^{(n)}_{\beta}={\rm{tr}}_{\,\ocH_{n'\setminus n}}R^{(n')}_\beta.
\eqno{(2.4.1)}$$

A consistent family of RDMs $R^{(n)}_\beta$ defines a state of
(i.e., a linear positive normalized functional on) the quasilocal
C$^*$-algebra constructed as the closure of the inductive limit of
$\fB_N$ as $N\to\infty$ where $\fB_N$ is the C$^*$-algebra of the bounded
operators in $\cH_N$, cf. Bratteli et al. (2002).
This motivates a study of properties of limiting RDM families
$\{R^{(n)}_\beta\}$. Our results in this direction are summarised
in Theorems \ref{main1} and \ref{main2}.
\medskip

\begin{theorem}\label{main1}
Fix $\beta >0$.
For $\sf P$-a.a. CDLT $T\in\cT_\infty$, $\forall$ $n=0,1,2,\ldots$, the family
of the RDMs $\{R^{(n)}_{\beta ,N},\;N=1,2,\ldots\}$ is compact in the trace
norm in $\cH_{n}$. Similarly, $\{R^{(n)}_{\beta ,N |\,\bx^{\rm c}(N)},\;N=1,2,\ldots\}$
is a compact family $\forall$ choice of the boundary conditions $\bx^{\rm c}(N)$.
\end{theorem}
\medskip

\begin{theorem}\label{main2}
Let $R^{(n)}_\beta$ be any limiting-point operator for the family
$\{R^{(n)}_{\beta ,N |\,\bx^{\rm c}(N)},$ $\;N=1,2,\ldots\}$. Then, $\forall$
${\tg}\in{\tG}$, operator $S^{(n)}({\tg})$ commutes with $R^{(n)}_{\beta}$:
$$R^{(n)}_{\beta}=S^{(n)}({\tg})R^{(n)}_{\beta}\left(S^{(n)}({\tg})\right)^{-1}.\eqno{(2.4.2)}$$
\end{theorem}
\medskip

\begin{remark} The statement of Theorem \ref{main2} is straightforward for
the limit points $R^{(n)}_{\beta}$ of the family $\{R^{(n)}_{\beta ,N},\;N=1,2,\ldots\}$
but requires a proof for the family $\{R^{(n)}_{\beta ,N |\,\bx^{\rm c}(N)},\;N=1,2,\ldots\}.$
\end{remark}
\medskip

The main role in the proof of Theorems \ref{main1} and \ref{main2} is played
by the Feynman--Kac (FK) representation for the RDMs $R^{(n)}_{\beta ,N}$
and $R^{(n)}_{\beta ,N |\,\bx^{\rm c}(N)}$ and their limiting counterparts
$R^{(n)}_\beta$. This representation is discussed in the next section.
\newpage

\section{The FK ensembles of paths and loops}
\vskip 1 truecm

{\bf 3.1. The FK representation for the Gibbs operators.} The Gibbs
operators $G_{\beta ,N}$ and $G_{\beta ,N |\,\bx^{\rm c}(N)}$
act as integral operators, with kernels $K_{\beta ,N}$ and
$K_{\beta ,N |\,\bx^{\rm c}(N)}$
$$\begin{array}{c}
\diy\big(G_{\beta ,N}\phi_N\big)(\bx (N) )
=\int\limits_{M^{\times\cV_N}}
{K}_{\beta ,N}(\bx (N),\by (N))
\phi_N(\by(N)){\rd}\by (N),\\
\diy\big(G_{\beta ,N|\,\bx^{\rm c}(N)}\phi_N\big)(\bx(N) )
=\int\limits_{M^{\times\cV_N}}
{K}_{\beta ,N|\,\bx^{\rm c}(N)}(\bx(N),\by(N))
\phi_N(\by (N)){\rd}\by(N).\end{array}\eqno (3.1.1)$$
Here $\by (N)=\{y(i):i\in\cV_N\}$ and we use a shorthand notation
${\rd}\by (N)=\prod\limits_{i\in\cV_N}v({\rd}y(i))$.

Further, the kernels $K_{\beta ,N}$ and $K_{\beta ,N |\,\bx^{\rm c}(N)}$
admit the FK-representations summarized in Lemma \ref{lem:3.1}.
The proof of this lemma follows the standard lines and is omitted.
The reader can confer Ginibre (1973) for details.

Given points $x,y\in M$, let $W^\beta_{x,y}$
denote the space of continuous
paths $\oom =\oom_{x,y}:\tau\in [0,\beta ]\mapsto\oom (\tau)\in M$,
of time-length $\beta$, beginning at
$x$ and terminating at $y$. Next, let $\bbP^\beta_{x,y}$ stand for the
(unnormalized) Wiener measure on $W^\beta_{x,y}$, with
$\bbP^\beta_{x,y}(W^\beta_{x,y})=p^\beta (x,y)$ where $p^\beta (x,y)$
is the value of the transition density from $x$ to $y$ in time $\beta$.
Furthermore, given particle configurations $\bx (N)=\{x(i)\},\by (N)=\{y(i)\}
\in M^{\times\cV_N}$, we set:
$$W^\beta_{\bx (N),\by (N)}=\operatornamewithlimits{\times}\limits_{i\in\cV_N}
W^\beta_{x(i),y(i)},\;\;
\bbP^\beta_{\bx (N),\by (N)}=\operatornamewithlimits{\times}\limits_{i\in\cV_N}
\bbP^\beta_{x(i),y(i)}. \eqno (3.1.2)$$
In other words, an element $\obom (N)=\oom_{\bx (N),\by (N)}
\in W^\beta_{\bx (N),\by (N)}$
is represented by a collection of paths $\{\oom_{x(i),y(i)}\}$ where
$\oom_{x(i),y(i)}\in W^\beta_{x(i),y(i)}$. We call such a collection a path
configuration over $\cV_N$. Moreover, under measure
$\bbP^\beta_{\bx (N),\by (N)}$ the paths $\oom_{x(i),y(i)}$ are independent
and each of them follows its own marginal measure $\bbP^\beta_{x(i),y(i)}$.

Further, we need to introduce functionals  $h(\obom (N))$ and
$h(\obom (N)|\,\bx^{\rm c} (N))$ describing an integral energy of
the path configuration $\oom(N)$ and its energy in the potential
field generated by $\bx^{\rm c} (N)$:
$$h(\obom (N))=\sum\limits_{(i,i')\in\cV_N\times\cV_N}
h^{i,i'}(\oom (i),\oom (i'))\eqno{(3.1.3)}$$
where $h^{i,i'}(\oom (i),\oom (i'))$ represents an integral along
trajectories $\oom (i)$ and $\oom (i')$. Namely, for $i=i'$
and $\oom\in W^\beta_{x(i),y(i)}$:
$$h^{i,i}(\oom ,\oom )=
\int\limits_0^{\beta}{\rd}\tau
\,U\big(\oom (\tau)\big).\eqno{(3.1.4)}$$
and for $i\neq i'$ and $\oom\in W^\beta_{x(i),y(i)}$, $\oom'\in W^\beta_{x(i'),y(i')}$:
$$h^{i,i'}(\oom ,\oom')=J({\td}(i,i'))
\int\limits_0^{\beta}{\rd}\tau
\,V\big(\oom (\tau),\oom' (\tau)\big).\eqno{(3.1.5)}$$
Pictorially, $h^{i,i}(\oom (i),\oom (i))$ yields an energy of the path
$\oom (i)$ in the external field generated by the potential $U$ and
$h^{i,i'}(\oom (i),\oom (i'))$ the
energy of interaction between paths $\oom (i)$ and $\oom (i')$.
Accordingly, $h(\obom (N))$ gives a full potential energy
of the path configuration $\obom (N)$.

Similarly,
$$h(\obom (N)|\,\bx^{\rm c} (N))=h(\obom (N))+\sum\limits_{i\in\cV_N, i'\in\oV_N}
h^{i,i'}(\oom (i),\bx^{\rm c}(i'))\eqno{(3.1.6)}$$
where
$h^{i,i'}(\oom (i),\bx^{\rm c}(i'))=J({\td}(i,i'))\int\limits_0^{\beta}{\rd}\tau
\,V\big(\oom (i,\tau),\bx^{\rm c}(i')\big)$.
\medskip

\begin{lemma}\label{lem:3.1} The integral kernels $K_{\beta ,N}
(\bx (N),\by (N))$
and\\ $K_{\beta ,N|\,\bx^{\rm c} (N)}(\bx (N),\by (N))$ are given by:
$$K_{\beta ,N}(\bx (N),\by (N))
=\diy\int\limits_{W^\beta_{\bx (N),\by (N)}}
{\bbP}^{\beta}_{\bx (N),\by (N)}
({\rd}\obom (N))
\exp\,\big[-h(\obom (N))\big]\eqno{(3.1.7)}$$
and
$$K_{\beta ,N|\,\bx^{\rm c} (N)}(\bx (N),\by (N))
=\diy\int\limits_{W^\beta_{\bx (N),\by (N)}}
{\bbP}^{\beta}_{\bx (N),\by (N)}
({\rd}\obom (N))
\exp\,\big[-h(\obom (N)|\,\bx^{\rm c} (N))\big]\eqno{(3.1.8)}$$
\end{lemma}
\medskip

{\bf 3.2. The FK representation for the partition functions and RDMs.}
Lemma \ref{lem:3.1} implies a working representation for
the partition functions $\Xi_{\beta ,N}$
and $\Xi_{\beta ,N|\,\bx^{\rm c}(N)}$ (see (2.2.5)). More precisely,
a key ingredient in the corresponding formulas will be
the space $W^\beta_{x,x}=W^\beta_x$ of closed paths
(starting and ending up at the same marked point $x\in M$); we
will employ the term ``loop" to make a distinction with a general case.
Accordingly, the notation $\bbP^\beta_{x,x}=\bbP^\beta_{x}$
will be in place here. Note that measure $\bbP^\beta_{x}$ in essence
does not depend on the choice of the point $x\in M$. Furthermore,
the notation $\om =\om_x\in W^\beta_x$
will be used for a loop with the marked initial/end point $x$,
omitting the bar in the previous symbol $\oom$. Next, we set:
$$W^\beta_{\bx (N)}=\operatornamewithlimits{\times}\limits_{i\in\cV_N}
W^\beta_{x(i)},\;\;
\bbP^\beta_{\bx (N)}=\operatornamewithlimits{\times}\limits_{i\in\cV_N}
\bbP^\beta_{x(i)}. \eqno (3.2.1)$$
An element $\bom (N)\in W^\beta_{\bx (N)}$
is represented by a collection of loops $\{\om_{x(i)}\}$ where
$\om_{x(i)}\in W^\beta_{x(i)}$; such a collection is called a loop
configuration over $\cV_N$. As before, under measure
$\bbP^\beta_{\bx (N)}$ the loops $\om_{x(i)}$ are independent
and each of them follows its own marginal measure $\bbP^\beta_{x(i)}$.

At this point we apply the Mercer theorem guaranteeing that the traces
${\rm{tr}}_{\cH_N}G_{\beta ,N}$ and
$\Xi_{\beta ,N|\,\bx^{\rm c}(N)}={\rm{tr}}_{\cH_N}G_{\beta ,N |\,\bx^{\rm c}(N)}$
are given by the integrals of the corresponding kernels
$K_{\beta ,N}(\bx (N),\by (N))$
and $K_{\beta ,N|\,\bx^{\rm c} (N)}(\bx (N),\by (N))$ along the diagonal
$\bx (N)=\by (N)$. This leads to Lemma \ref{lem:3.2} below.

Let us denote:
$$\int{\rd}\bom (N):=\int_{M^{\times\cV_N}}
{\rd}\bx (N)\int\limits_{W^\beta_{\bx (N)}}
{\bbP}^{\beta}_{\bx (N)}({\rd}\bom (N)).\eqno (3.2.2)$$
\medskip

\begin{lemma}\label{lem:3.2} The partition functions $\Xi_{\beta ,N}$
and $\Xi_{\beta ,N|\,\bx^{\rm c} (N)}$ are given by:
$$\Xi_{\beta ,N} =\diy
\int{\rd}\bom (N)\exp\,\big[-h(\bom (N))\big]\eqno{(3.2.3)}$$
and
$$\Xi_{\beta ,N|\,\bx^{\rm c} (N)}=\diy\int{\rd}\bom (N)
\exp\,\big[-h(\bom (N)|\,\bx^{\rm c} (N))\big]\,.\eqno{(3.2.4)}$$
\end{lemma}
\medskip

Let us now turn to the RDMs $R^{(n)}_{\beta ,N}$ and
$R^{(n)}_{\beta ,N|\,\bx^{\rm c} (N)}$. These operators are again given
by their integral kernels:
$$\begin{array}{c}
\diy\big(R^{(n)}_{\beta ,N}\phi_n\big)(\bx (n) )
=\int\limits_{M^{\times\cV_n}}
F^{(n)}_{\beta ,N}(\bx (n),\by (n))
\phi_n(\by (n)){\rd}\by (n),\\
\diy\big(R^{(n)}_{\beta ,N|\,\bx^{\rm c}(N)}\phi_n\big)(\bx (n))
=\int\limits_{M^{\times\cV_n}}
F^{(n)}_{\beta ,N|\,\bx^{\rm c}(N)}(\bx(n),\by(n))
\phi_n(\by (n)){\rd}\by (n).\end{array}\eqno (3.2.5)$$
Next, $F^{(n)}_{\beta ,N}(\bx (n),\by (n))$ and
$F^{(n)}_{\beta ,N|\,\bx^{\rm c}(N)}(\bx(n),\by(n))$
are called reduced density matrix kernels (RDMKs). They can be written in the form
$$\begin{array}{c}
\diy F^{(n)}_{\beta ,N}(\bx (n),\by (n))=
\frac{\Xi^{(n)}_{\beta ,N}(\bx (n),\by (n))}{\Xi_{\beta ,N}},\\
\diy F^{(n)}_{\beta ,N|\,\bx^{\rm c}(N)}(\bx(n),\by(n))=
\frac{\Xi^{(n)}_{\beta ,N|\,\bx^{\rm c} (N)}(\bx (n),\by (n))
}{\Xi_{\beta ,N|\,\bx^{\rm c} (N)}}\end{array}\eqno (3.2.6)$$
where quantities $\Xi^{(n)}_{\beta ,N}(\bx (n),\by (n))$ and
$\Xi^{(n)}_{\beta ,N|\,\bx^{\rm c} (N)}(\bx (n),\by (n))$ admit\\
representations similar to (3.2.3) and (3.2.4), see Lemma
\ref{lem:3.3}.

We will use a notation similar to (3.2.2):
$$\int{\rd}\bom (N\setminus n):=\int_{M^{\times\cV_N\setminus\cV_n}}
{\rd}\bx (N\setminus n)\int\limits_{W^\beta_{\bx (N\setminus n)}}
{\bbP}^{\beta}_{\bx (N\setminus n)}({\rd}\bom (N\setminus n))
\eqno (3.2.7)$$
where $\bx (N\setminus n)$ stands for a particle configuration
$\{x(j),j\in\cV_N\setminus\cV_n\}$ and $\bom (N\setminus n)$ for the loop
configuration $\{\om (j),j\in\cV_N\setminus\cV_n\}$. Symbol $\vee$ will
be used for concatenation of particle configurations and for concatenation
of path and loop configurations (originally
defined over disjoint sets). Accordingly, for a path configuration
$\obom (n)\in W^\beta_{\bx (n),\by (n)}$ over $\cV_n$ and a
loop configuration $\bom (N\setminus n)$ over $\cV_N\setminus\cV_n$,
the energies $h(\obom (n)\vee\bom (N\setminus n))$ and
$h(\obom (n)\vee\bom (N\setminus n)|\,\bx^{\rm c} (N))$ are defined as in
(3.1.2)--(3.1.5).
\medskip

\begin{lemma}\label{lem:3.3}
The numerators $\Xi^{(n)}_{\beta ,N}(\bx (n),\by (n))$ and \\
$\Xi^{(n)}_{\beta ,N|\,\bx^{\rm c} (N)}(\bx (n),\by (n))$
are given by:
$$\begin{array}{r}
\diy\Xi^{(n)}_{\beta ,N}(\bx (n),\by (n))=
\diy\int\limits_{W^\beta_{\bx (n),\by (n)}}
{\bbP}^{\beta}_{\bx (n),\by (n)}({\rd}\obom (n))\qquad{}\\
\diy\times\int{\rd}\bom (N\setminus n)
\exp\,\big[-h(\obom (n)\vee\bom (N\setminus n))\big]
\end{array}\eqno{(3.2.8)}$$
and
$$\begin{array}{r}
\diy\Xi^{(n)}_{\beta ,N|\,\bx^{\rm c} (N)}(\bx (n),\by (n))=
\diy\int\limits_{W^\beta_{\bx (n),\by (n)}}
{\bbP}^{\beta}_{\bx (n),\by (n)}({\rd}\obom (n))\qquad{}\\
\diy\times\int{\rd}\bom (N\setminus n)
\exp\,\big[-h(\obom (n)\vee\bom (N\setminus n)|\,\bx^{\rm c} (N))\big]
\,.\end{array}\eqno{(3.2.9)}$$
\end{lemma}
\medskip

The proof of Lemma \ref{lem:3.3} consists in translating the
partial traces into the integrals of the kernels
$F^{(n)}_{\beta ,N}$ and $F^{(n)}_{\beta ,N|\,\bx^{\rm c}(N)}$
of the operators $R^{(n)}_{\beta ,N}$ and
$R^{(n)}_{\beta ,N|\,\bx^{\rm c}(N)}$. We omit it from the paper.
\medskip

{\bf 3.3. The FK-DLR equations.}
The representations (3.2.3)--(3.2.4) suggest introducing
probability distributions $\mu_N$ and $\mu_{N|\,\bx^{\rm c} (N)}$
on loop configurations $\bom_N$,
with the densities (the Radon--Nikodym derivatives)
$$\begin{array}{c}
\diy p_N(\bom (N)):=\frac{\mu_N({\rd}\bom (N))}{{\rd}\bom (N)}=
\frac{\exp\,\big[-h(\bom (N))\big]}{\Xi_{\beta ,N}}\\
\diy p_{N|\,\bx^{\rm c} (N)}(\bom (N)):=\frac{\mu_{N|\,\bx^{\rm c} (N)}({\rd}\bom (N))}{{\rd}\bom (N)}=
\frac{\exp\,\big[-h(\bom (N)|\,\bx^{\rm c} (N))\big]}{\Xi_{\beta ,N|\,\bx^{\rm c} (N)}}
\end{array}\eqno{(3.3.1)}$$
A crucial property is that the measures $\mu_N$ and $\mu_{N|\,\bx^{\rm c} (N)}$
satisfy DLR (Dobrushin-Lanford-Ruelle)-type equations. Namely,
let $p^{(n)}_N(\bom(n)|\bom (N\setminus n))$ and \\
$p^{(n)}_{N|\,\bx^{\rm c} (N)}(\bom (n)|\bom (N\setminus n))$ stand for the conditional
densities generated by $\mu_N$ and $\mu_{N|\,\bx^{\rm c} (N)}$, respectively, for
the loop configuration $\bom (n)$ over $\cV_n$ given a loop configuration
$\bom (N\setminus n)$ over $\cV_N\setminus\cV_n$. Then
$$\begin{array}{c}
\diy p^{(n)}_N(\bom (n)|\bom (N\setminus n))
:=\frac{\mu_N({\rd}\bom (n)|\bom (N\setminus n))}{{\rd}\bom (N)}=
\frac{\exp\,\big[-h(\bom (n)|\bom (N\setminus n))\big]}{\Xi_{\beta ,n}(\bom (N\setminus n))}\\
\begin{array}{r}\diy p^{(n)}_{N|\,\bx^{\rm c} (n)}(\bom (n)|\bom (N\setminus n)):
=\frac{\mu_{N|\,\bx^{\rm c} (N)}({\rd}\bom (n)|\bom (N\setminus n))}{{\rd}\bom (N)}\qquad\qquad{}\\
\diy =
\frac{\exp\,\big[-h(\bom (n)|\,\bom (N\setminus n)\vee\bx^{\rm c} (N))\big]}{\Xi_{\beta ,n|\,\bx^{\rm c} (N)}
(\bom (N\setminus n))}\end{array}
\end{array}\eqno{(3.3.2)}$$
Here $h(\bom (n)|\bom (N\setminus n))$ and $h(\bom (n)|\,\bom (N\setminus n)\vee\bx^{\rm c} (N))$
stand for `conditional' energies and $\Xi_{\beta ,n}(\bom (N\setminus n))$
and $\Xi_{\beta ,n|\,\bx^{\rm c} (N)}(\bom (N\setminus n))$ for `conditional' partition
functions:
$$h(\bom (n)|\bom (N\setminus n))=h(\bom (n)\vee \bom (N\setminus n))
-h(\bom (N\setminus n)),\eqno{(3.3.3)}$$
$$\begin{array}{r}h(\bom (n)|\bom (N\setminus n)\vee\bx^{\rm c} (N))=
h(\bom (n)\vee\bom (N\setminus n)|\,\bx^{\rm c} (N))\qquad{}\\
-h(\bom (N\setminus n)|\,\bx^{\rm c} (N)),\end{array}
\eqno{(3.3.4)}$$
$$\Xi_{\beta ,n}(\bom (N\setminus n)) =\diy
\int{\rd}\bom (n)\exp\,\big[-h(\bom (n))|\,\bom (N\setminus n)\big]
\eqno{(3.3.5)}$$
and
$$\Xi_{\beta ,n|\,\bx^{\rm c} (N)}(\bom (N\setminus n))=\diy\int{\rd}\bom (n)
\exp\,\big[-h(\bom (n)|\,\bom (N\setminus n)\vee\bx^{\rm c} (N))\big]\,.
\eqno{(3.3.6)}$$

We call Eqn (3.3.2) the FK-DLR equation in volume $\cV_N$.

Concluding this section, we give an expression for the kernels
$F^{(n)}_{\beta ,N}(\bx (n),\by (n))$ and $F^{(n)}_{\beta ,N|\,\bx^{\rm c}(N)}(\bx(n),\by(n))$:
$\forall$ $0< n\leq n'<N$:
$$\begin{array}{l}
\diy F^{(n)}_{\beta ,N}(\bx (n),\by (n))=\int{\rd}\bom (N\setminus n')
\frac{p_N^{(N\setminus n')}(\bom (N\setminus n'))}{\Xi_{\beta ,n'}(\bom (N\setminus n'))}
\int{\rd}\bom (n'\setminus n)\\
\diy\quad\times\int_{W^\beta_{\bx (n),\by (n)}}\bbP^\beta_{\bx (n),\by (n)}({\rd}\obom (n))
\exp\,[-h(\obom (n)\vee\bom (n'\setminus n)|\,\bom (N\setminus n'))],\\
\diy F^{(n)}_{\beta ,N|\,\bx^{\rm c}(N)}(\bx(n),\by(n))=
\int{\rd}\bom (N\setminus n')
\frac{p_{N|\,\bx^{\rm c}(N)}^{(N\setminus n')}(\bom (N\setminus n'))}{
\Xi_{\beta ,n'}(\bom (N\setminus n')\vee\bx^{\rm c} (N))}\\
\diy\quad\times\int{\rd}\bom (n'\setminus n)
\int_{W^\beta_{\bx (n),\by (n)}}\bbP^\beta_{\bx (n),\by (n)}({\rd}\obom (n))\\
\diy\qquad\times
\exp\,[-h(\obom (n)\vee\bom (n'\setminus n)|\,\bom (N\setminus n')\vee\bx^{\rm c} (N))].
\end{array}\eqno (3.3.7)$$
For $n=n'$, the integral $\diy\int{\rd}\bom (n'\setminus n)$ is omitted.

Our next goal is to write down FK-DLR equations for the whole of $\cV=\cV(T)$.
Here we consider a probability measure $\mu =\mu_{\cV}$ on infinite
loop configurations $\bOm =\bOm_{\cV}$ over $\cV $ (for a formal
background, see Kelbert and Suhov (2013, \cite{KS1})). The equation is written for $p^{(n)}(\bom (n)|\bOm^{\rm c} (n))$,
the conditional probability density for a loop configuration $\bom (n)$
over $\cV_n$, given a loop configuration $\bOm^{\rm c} (n)$ over $\cV\setminus\cV_n$.
This density should be given by
$$\begin{array}{c}
\diy p^{(n)}(\bom (n)|\bOm^{\rm c} (n))
:=\frac{\mu ({\rd}\bom (n)|\bOm^{\rm c} (n))}{{\rd}\bom (n)}=
\frac{\exp\,\big[-h(\bom (n)|\bOm^{\rm c} (n))\big]}{\Xi_{\beta ,n}(\bOm^{\rm c} (n))}\,.
\end{array}\eqno{(3.3.8)}$$

Like $h(\bom (n)|\bom (N\setminus n))$ and $\Xi_{\beta ,n}(\bom (N\setminus n))$
before, the quantities $h(\bom (n)|\bOm^{\rm c} (n))$ and $\Xi_{\beta ,n}(\bOm^{\rm c} (n))$
represent the conditional energy and the conditional partition function. They
can be defined as the limits
$$h(\bom (n)|\bOm^{\rm c} (n))=\lim_{N\to\infty}
h(\bom (n)|\bOm (N\setminus n)),\eqno{(3.3.9)}$$
$$\Xi_{\beta ,n}(\bOm^{\rm c} (n))=\lim_{N\to\infty}\Xi_{\beta ,n}(\bOm (N\setminus n))
\eqno{(3.3.10)}$$
where $\bOm (N\setminus n)$ stands for the restriction
of $\bOm^{\rm c} (n)$ to $\cV_N\setminus\cV_n$. The existence of the limit will
be guaranteed by the assumption (2.3.5) $\forall$ $\bom (n)$ and $\bOm^c (n)$
for ${\sf P}$-a.a. $T\in{\mathcal T}_\infty$.

Formulas (3.3.7) admit a generalization to the infinite-volume situation:
$\forall$ $0\leq n<n'$,
$$\begin{array}{l}\diy F^{(n)}_\beta(\bx (n),\by (n))=\int
\frac{\mu({\rd}\bOm^{\rm c} (n'))}{\Xi_{\beta ,n'}(\bOm^{\rm c} (n'))}
\int{\rd}\bom (n'\setminus n)\\
\diy\quad\times\int_{W^\beta_{\bx (n),\by (n)}}\bbP^\beta_{\bx (n),\by (n)}
({\rd}\obom (n))
\exp\,[-h(\obom (n)\vee\bom (n'\setminus n)|\,\bOm^{\rm c} (n'))];\end{array}
\eqno (3.3.11)$$
owing to the FK-DLR propety, the RHS in (3.3.11) does not depend on the choice of $n'>n$.
Moreover, the integral $$\diy\int{\rd}\bx (n)F^{(n)}_\beta(\bx (n),\bx (n))=\mu (\cV )=1.$$

Consider the operator $R^{(n)}_\beta$ in $\cH(n)={\cH}^{\otimes \cV_n}$ with the integral kernel
$F^{(n)}_\beta(\bx (n),\by (n))$ given by (3.3.4). The aforementioned properties
imply that the trace ${\rm{tr}}_{\cH (n)}R^{(n)}_\beta =1$ and the following
compatibility relation holds true:
$$R^{(n)}_\beta={\rm{tr}}_{\cH (n'\setminus n)}R^{(n')}_\beta .\eqno (3.3.12)$$
Thus, were the operators $R^{(n)}_\beta$ positive definite, we could
speak of an infinite-volume state of the quasilocal C$^*$-algebra $\fB$.
Cf. Remark 2.2. Notwithstanding, we state our main result:
\vskip .5 truecm

\begin{theorem}\label{Main1}
Under the above assumptions, any limit-point operator $R^{(n)}_\beta$
from Theorem \ref{main1} is a positive definite trace-class integral
operator of trace $1$ and with the kernel
$F^{(n)}_\beta$ admitting the representation (3.3.4) where
probability distribution $\mu$ satisfies the infinite-volume FK-DLR
equations (3.3.8).
\end{theorem}

\begin{theorem}\label{Main2}
Let an integral operator $R^{(n)}_\beta$ admit
the representation (3.3.11) where
probability distribution $\mu$ satisfies the infinite-volume FK-DLR
equations (3.3.8). Then $\forall$ ${\tg}\in{\tG}$
$$
R^{(n)}_\beta=S^{(n)}({\tg}) R^{(n)}_\beta (S^{(n)}({\tg}))^{-1}.
\eqno {(3.3.13)}$$

\end{theorem}
\newpage

\section{The proofs: the compactness and the tuned-action arguments}

The proof of Theorems \ref{main1} and \ref{Main1} is based on
a compactness argument (cf. Kelbert and Suhov (2013, \cite{KS1} and \cite{KS2}). We want
to note that this argument does not depend upon the
dimensionality of the system.
\vskip .5 truecm

{\bf 4.1. Proof of Theorems \ref{main1} and \ref{Main1}.} As
in Kelbert and Suhov (2013, \cite{KS1}, \cite{KS2}), we first prove that, $\forall$
$n\geq 0$, the sequences of RDMKs
$\{F^{(n)}_{\beta ,N},\;N=n+1,n+2,\ldots\}$ and
$\{F^{(n)}_{\beta ,N |\,\bx^{\rm c}(N)},\;N=n+1,n+2,\ldots\}$
are compact in the space $C^0\left(M^{\times\cV_n}\times M^{\times\cV_n}\right)$.
Applying Lemma 1.5 from Kelbert and Suhov (2013, \cite{KS1}) (this lemma goes back to Suhov (1970)),
we will obtain that the sequences of RDMs
$\{R^{(n)}_{\beta ,N}\}$ and
$\{R^{(n)}_{\beta ,N |\,\bx^{\rm c}(N)}\}$
are compact in the trace-norm topology in $\cH(n)$. This yields the
statement of Theorem \ref{main1}. A straightforward
consequence of the convergence will be that
any limiting RDMK $F^{(n)}_\beta$ admits the representation (3.3.4)
where $\mu$ satisfies the infinite-volume FK-DLR equation, i.e.,
the assertion of Theorem \ref{Main1}.

To verify compactness of the RDMKs $\{F^{(n)}_{\beta ,N}\}$ and
$\left\{F^{(n)}_{\beta ,N |\,\bx^{\rm c}(N)}\right\}$, we follow the same line
as in Kelbert and Suhov (2013, \cite{KS1} and \cite{KS2}), i.e., employ the Ascoli--Arzela theorem.
To this end, we need to check the properties of uniform boundedness
and equicontinuity. For definiteness, we focus on the (slightly
more complex) case of the
sequence $\{F^{(n)}_{\beta ,N |\,\bx^{\rm c}(N)}\}$.

More precisely, to show uniform boundedness, we first use an upper
bound for the number of vertices $k_i$ on $\cV_i\setminus\cV_{i-1}$
under the measure $\sf P$; cf. Kelbert et al. (2013, \cite{KSY}), Eqn (4.1). Namely, $\forall$
$\varepsilon\in (0,1)$,
for $\sf P$-a.a. $T\in{\mathcal T}_\infty$
$\exists$ a constant $C=C(T)$ such that
$$
k_i \le C i \big(\ln\;i\big)^{1/2+\varepsilon },\;\; i=2,3,\ldots ,
\eqno (4.1.1)$$
(see Kelbert et al. \cite{KSY}). This yields that
$$\sum\limits_{i=1}^{\infty}k_iJ(i)< C_1(T)+C(T)\sum\limits_{i=2}^{\infty}
i(\ln i)^{1/2+\varepsilon}\Big(\frac{1}{i\ln i}\Big)^3:=C(T)J^*.\eqno (4.1.2)$$

We use (4.1.1) and (4.1.2) to bound the quantity
$$\begin{array}{l}
\diy q(\obom (n)|\,\bom (N\setminus n)\vee\bx^{\rm c}(N)) :=\\
\diy\qquad \frac{\exp\,\Big[-h(\obom (n)|
\,\bom (N\setminus n)\vee\bx^{\rm c} (N))\Big]}{\Xi_{\beta ,n}
\big(\bom (N\setminus n)\vee\bx^{\rm c} (N)\big)};
\end{array}\eqno (4.1.3)$$
cf. (3.3.7) for $n'=n$. Namely, (4.1.2) implies that, for ${\sf P}$-a.a.
$T\in{\mathcal T}_\infty$, $\forall$ $n\geq 0$ and $N>n$,
$$\begin{array}{l}\diy
\exp\left[-\beta ({\ov U}+C(T)J^*{\ov V})\sharp\,\cV_n\right]\\
\diy\quad\leq \exp\,\Big[-h(\obom (n)
|\,\bom (N\setminus n)\vee\bx^{\rm c} (N))\Big]\\
\diy\qquad\qquad\qquad\qquad\qquad\qquad
\leq \exp\left[\beta ({\ov U}+C(T)J^*{\ov V})\sharp\,\cV_n\right]
\end{array}\eqno (4.1.4)$$
for all path configurations $\obom\in W_{\bx (n),\by (n)}$ and loop configurations
$\bom (N\setminus n)\in W_{\cV_N\setminus\cV_n}$. Here $\sharp\,\cV_n = \sum_{i=1}^{n} k_i$ stands
for the number of vertices in the set  $\cV_n$, cf. (4.1.1).

The lower bound in (4.1.4) yields that
$$\Xi_{\beta ,n}
\big(\bom (N\setminus n)\vee\bx^{\rm c} (N)\big)\geq
\exp\left[-\beta ({\ov U}+C(T)J^*{\ov V})\sharp\,\cV_n\right]
\times \big({\ovp}^\beta_M\big)^{\sharp\,\cV_n},\eqno (4.1.5)$$
where
$${\ovp}^\beta_M=\frac{1}{(2\pi\beta )^{d/2}}
\sum\limits_{\unn =(n_1,\ldots ,n_d)\in{\bbZ}^d}\exp\left(-|\unn |^2\big/2\beta\right)
\eqno (4.1.6)$$
is the probability density of transition from $x\in M$ to $x$ in time $\beta$
in the Brownian motion on $M$.
Next, (4.1.5) and the upper bound in (4.1.4) imply that
$$
q(\obom (n)|\bom (N\setminus n)\vee\bx^{\rm c}(N))
\leq\frac{1}{\big({\ovp}^\beta_M\big)^{\sharp\,\cV_n}}
\exp\left[2\beta ({\ov U}+C(T)J^*{\ov V})\sharp\,\cV_n\right].
\eqno (4.1.7)$$
Substituting (4.1.7) in (3.3.7), we obtain that
$$F^{(n)}_{\beta ,N |\,\bx^{\rm c}(N)} (\bx (n),\by (n))\leq
\exp\left[2\beta ({\ov U}+C(T)J^*{\ov V})\sharp\,\cV_n\right]\eqno (4.1.8)
$$
which gives the desired uniform upper bound.

To check equicontinuity, we analyze the derivatives \\
$\nabla_{x(i)}F^{(n)}_{\beta ,N |\,\bx^{\rm c}(N)} (\bx (n),\by (n))$ and
$\nabla_{y(i)}F^{(n)}_{\beta ,N |\,\bx^{\rm c}(N)} (\bx (n),\by (n))$, $i\in\cV_n$.
Again we use the representation (3.3.7) with $n=n'$.
We need to differentiate the integral
$$\int\limits_{W^\beta_{\bx (n),\by (n)}}\bbP^\beta_{\bx (n),\by (n)}({\rd}\obom (n))
\exp\,[-h(\obom (n)|\,\bom (N\setminus n)\vee\bx^{\rm c} (N))].
\eqno (4.1.9)$$
For definiteness, consider one of the gradients $\nabla_{y(i)}$. It is convenient to
represent the integral (4.1.8) in the form
$$\begin{array}{l}\diy\prod\limits_{j\in\cV_n}p^\beta_M(x(j),y(j))\frac{1}{\left[{\ovp}^\beta_M\right]^{\#\cV_n}}
\int\limits_{W^\beta_{\bx (n),\bx (n)}}\bbP^\beta_{\bx (n),\bx (n)}({\rd}\bom (n))\\
\diy\qquad\times\exp\,\Big\{-h\Big[\big(\bom (n)+\oboeta (n)\big)
|\,\bom (N\setminus n)\vee\bx^{\rm c} (N)\Big]\Big\}\,.\end{array}
\eqno (4.1.10)$$
Here $p^\beta_M(x,y)$ denotes the transition probability density
$$p^\beta_M(x,y)=\frac{1}{(2\pi\beta )^{d/2}}
\sum\limits_{\unn =(n_1,\ldots ,n_d)\in{\bbZ}^d}\exp\left(-|x-y
+\unn |^2\big/2\beta\right)\,,\eqno (4.1.11)$$
and ${\ovp}^\beta_M$ has been determined in (4.1.6).

Next, $\oboeta (n)=\{\eta (j),j\in\cV_n\}$ is a collection of linear paths
$$\eta (j,\tau )=\frac{\tau}{\beta}(y(j)-x(j)),\;\;j\in\cV_n,\eqno (4.1.12)$$
and the component-wise addition in $\bom (n)+\oboeta (n)$:
$$\begin{array}{l}\bom (n)+\oboeta (n)=\{\om (j)+\eta (j),\;j\in\cV_n\}\\
\quad\hbox{where }
\om (j)+\eta (j):\;\tau\in [0,\beta ]\mapsto\big(\om (j,\tau )+\eta (j,\tau )\big)\,
{\rm{mod}}\;1.\end{array}\eqno (4.1.13)$$
It is now clear that there will be two contributions into\\
$\nabla_{y(i)}F^{(n)}_{\beta ,N |\,\bx^{\rm c}(N)} (\bx (n),\by (n))$: one
coming from
$$\nabla_{y(i)}p^\beta_M(x(j),y(j)),
$$
the other from
$$\begin{array}{l}
\nabla_{y(i)}\exp\,\Big\{-h\Big[\big(\bom (n)+\oboeta (n)\big)
|\,\bom (N\setminus n)\vee\bx^{\rm c} (N)\Big]\Big\}\\
\diy\quad =-\nabla_{y(i)}h\Big[\big(\bom (n)+\oboeta (n)\big)
|\,\bom (N\setminus n)\vee\bx^{\rm c} (N)\Big]\\
\diy\qquad\times\exp\,\Big\{-h\Big[\big(\bom (n)+\oboeta (n)\big)
|\,\bom (N\setminus n)\vee\bx^{\rm c} (N)\Big]\Big\}\,.
\end{array}$$
The uniform bound
$$\left|\nabla_{y(i)}p^\beta_M(x(j),y(j))
\right|\leq C(\beta )\in (0,+\infty )\eqno (4.1.14)$$
is straightforward. Next, we have the estimate
$$\begin{array}{r}
\left|\nabla_{y(i)}h\Big[\big(\bom (n)+\oboeta (n)\big)
|\,\bom (N\setminus n)\vee\bx^{\rm c} (N)\Big]\right|\quad{}\\
\leq\beta\big(\#\cV_n\big)[{\ov U}+C(T)J^*{\ov V}].\end{array}$$
Together with (4.1.4) it implies that
$$\begin{array}{l}
\nabla_{y(i)}\exp\,\Big\{-h\Big[\big(\bom (n)+\oboeta (n)\big)
|\,\bom (N\setminus n)\vee\bx^{\rm c} (N)\Big]\Big\}\\
\diy\qquad\leq \big(\#\cV_n\big)[{\ov U}+C(T)J^*{\ov V}]
\exp\left[\beta ({\ov U}+C(T)J^*{\ov V})\sharp\,\cV_n\right].
\end{array}\eqno (4.1.15)$$
The bounds (4.1.14) and (4.1.15) lead to a uniform bound
upon\\ $\left|\nabla_{y(i)}F^{(n)}_{\beta ,N |\,\bx^{\rm c}(N)}
(\bx (n),\by (n))\right|$.
This completes the proof of Theorems \ref{main1} and \ref{Main1}.$\quad\Box$
\medskip

{\bf 4.2. Proof of Theorems \ref{main2} and \ref{Main2}.}
Theorem \ref{main2} follows from Theorem \ref{Main2}; therefore,
we focus on the proof of Theorem \ref{Main2}. Eqn (3.3.13) follows
from the property
$$\lim_{n'\to\infty}\frac{q_{n'}^{(n)}(S({\tt g})\obom (n)|\,\bOm^{\rm c} (n'))
}{q_{n'}^{(n)}(\obom (n)|\,\bOm^{\rm c} (n'))}=1,\;{\tt g}\in{\tt G},
\eqno (4.2.1)$$
uniformly in the path configurations $\obom (n)=\{\om (j),j\in\cV_n\}\in W^\beta_{\bx (n),\by (n)}$
and the loop configurations $\bOm^{\rm c} (n')$ over $\cV\setminus\cV_{n'}$.
Here, the functional $q_{n'}^{(n)}(\obom (n)|\,\bOm^{\rm c} (n'))$
emerges from representation (3.3.11):
$$\begin{array}{l}
\diy q_{n'}^{(n)}(\obom (n)|\,\bOm^{\rm c} (n'))(\bx(n),\by(n))=
\frac{1}{\Xi_{\beta ,n'}(\bOm^{\rm c} (n'))}
\int{\rd}\bom (n'\setminus n)\\
\diy\quad\times\int_{W^\beta_{\bx (n),\by (n)}}\bbP^\beta_{\bx (n),\by (n)}
({\rd}\obom (n))
\exp\,[-h(\obom (n)\vee\bom (n'\setminus n)|\,\bOm^{\rm c} (n'))]\,,\end{array}
\eqno (4.2.2)$$
and
$$\begin{array}{l}S({\tt g})\obom (n)=\{S({\tt g})\om (i),i\in\cV_n\}\\
\qquad\qquad\hbox{where}\;
S({\tt g})\om (i):\tau\in [0,\beta ]\mapsto S({\tt g})\om (i,\tau).\end{array}$$

To check (4.2.1), we again follow
the argument used in Kelbert and Suhov (2013, \cite{KS1}, \cite{KS2}) (which goes back to
Pfister (1981) and Fr\"ohlich and Pfister. (1981); cf. also Georgii (1988)). The backbone of the argument
is the following inequality: $\forall$ given $a>1$, ${\tt g}\in {\tt G}$
and positive integer $n$, if $n'$ is large enough
then, $\forall$ $\obom (n)\in W^\beta_{\bx (n),\by (n)}$, $\bx(n),\by(n)\in M^{\times\cV_n}$
and the loop configurations $\bOm^{\rm c} (n')$ over $\cV\setminus\cV_{n'}$,
$$\begin{array}{r}aq_{n'}^{(n)}({\tt g}\obom (n)|
\bOm^{\rm c}(n'))
+aq_{n'}^{(n)}({\tt g}^{-1}\obom (n)|
\bOm^{\rm c} (n'))\qquad{}\\
\geq 2 q_{n'}^{(n)}(\obom (n)|
\bOm^{\rm c}(n')).\end{array}\eqno (4.2.3)$$

The verification of Eqn (4.2.3) is based on a special construction
related to a family of ``tuned" actions $\bttg_{n'\setminus n} \bom(n'\setminus n)$
on loop configurations $\bom (n'\setminus n)$; see Eqns (4.2.5), (4.2.6) below. 
(A tuned action can be described as an interpolation between the unity (identity)
and the group action by $g$.)
A particular feature of the tuned action $\bttg_{n'\setminus n}$ is that
it ``decays" to ${\tt e}$, the unit element of ${\tt G}$ (which generates
a ``trivial" identity action), when we move the vertex of the tree $T$
from $\cV_n$ towards $\cV\setminus\cV_{n'}$.
Formally, (4.2.3) is implied by the following estimate: $\forall$ given $n$,
$\obom (n)\in\oW^{\beta}_{\bx (n), \by (n)}$, ${\ttg}\in{\ttG}$ and $a \in (1,\infty )$,
for any $n'$ large enough, $\bom (n'\setminus n)$ and
$\bOm^{\rm c}(n')$,
$$\begin{array}{l}\diy\frac{a}{2}\exp\,\Big[-h\bigl({\ttg}\obom (n)\vee
\bttg_{n'\setminus n}\bom (n'\setminus n) \ \bigl|\bigr. \
\bOm^{\rm c}(n')\bigr) \Big]\\
\qquad\quad\diy +
\frac{a}{2}\exp\,\Big[-h\bigl({\ttg}^{-1}\obom (n)\vee
\bttg^{-1}_{n'\setminus n}\bom (n'\setminus n)  \ \bigl|\bigr. \
\bOm^{\rm c}(n')\bigr)\Big]\\
\qquad\qquad\geq\diy
\exp\,\Big[-h\bigl(\obom (n)\vee\bom (n'\setminus n)  \ \bigl|\bigr. \
\bOm^{\rm c}(n')\bigr) \Big]\,.
\end{array}\eqno (4.2.4)$$

Indeed, (4.2.3) follows from (4.2.4) by integrating in ${\rd}\bom (n'
\setminus n)$ and normalizing
by $\Xi_{\beta, n'}(\bOm^{\rm c} (n'))$;
cf. (4.2.2). Here it is important that the Jacobian of the map
$\bom (n'\setminus n)\mapsto\bttg_{n'\setminus n}
\bom (n'\setminus n)$ is equal to $1$.

The rest of the argument concentrates on verifying (4.2.4). The tuned family
$\bttg_{n'\setminus n}$ consists of individual
actions ${\ttg}^{(n')}_j\in{\ttG}$ at vertices $j\in\cV_{n'}\setminus\cV_n$:
$$\bttg_{n'\setminus n}=\{{\ttg}^{(n')}_j,\;j\in\cV_{n'}
\setminus\cV_n\}.\eqno (4.2.5)$$
We use the representation (2.3.3) and identify the element
${\ttg}\in{\tt G}$ with a vector $\utheta =\theta A\in{\bbR}^d$.
Then the actions ${\ttg}^{(n')}_j\in {\ttG}$ correspond to multiples of
the vector $\utheta$; cf. Eqn (4.2.6) below. It is convenient to fix a positive
integer $\ovr>n$ and identify
$${\ttg}^{(n')}_j\;\hbox{ with \;\;}\utheta {\gamma (n',k)}\eqno (4.2.6)$$
where $k={\tt d}(s_0,j)$ (recall, $s_0$ is the root of $T$) and
$$\gamma (n',k)
=\begin{cases}1,&k\leq\ovr,\\
\vartheta\big(k-\ovr,n'-\ovr\big),&k>\ovr .
\end{cases}\eqno (4.2.7)$$
In turn, the function $\vartheta (a,b)$ is determined by
$$\vartheta (a,b)={\mathbf 1}(a\leq 0)+
\frac{{\mathbf 1}(0<a<b)}{Q(b)}\int_a^bz(u){\rd}u,\;\;
a,b\in\bbR ,\eqno (4.2.8)$$
with the same functions $Q(b)$ and $z(u)$ as in Fr\"ohlich and Pfister (1981)
$$\begin{array}{l}\diy Q(b)=\int_0^b z(u){\rd}u,\\
\diy\quad\hbox{where}
\;z(u)={\mathbf 1}(u\leq 2)+{\mathbf 1}(u>2)
\frac{1}{u\ln\,u},\;b>0.\end{array}\eqno (4.2.9)$$

Next, $\bttg_{n'\setminus n}^{-1}$ is the
collection of the inverse elements:
$$\bttg_{n'\setminus n}^{-1}=
\left\{{{\ttg}_j^{(n')}}^{-1},\;j\in\cV_{n'}\setminus\cV_n\right\}.$$
It will be convenient to use formulas (2.4.6)--(2.4.7) for ${\ttg}^{(n)}_j$
for $j\in\cV_n$, or even for $j\in\cV$,
as these formulas agree with the requirement that ${\ttg}^{(n')}_j\equiv{\ttg}$
when $j\in\cV_n$ and ${\ttg}^{(n')}_j\equiv{\tt e}$ for $j\in\cV
\setminus\cV_{n'}$. Accordingly, we will employ the notation $\bttg_{n'}
=\{{\ttg}^{(n')}_j,\;j\in\cV_{n'}\}$.

Next, we use the invariance property (2.3.4). The
Taylor formula for the function $V\in {\mathbf C}^2(\mathbb R^2)$ yields for
$j,j'\in\cV_n$:
$$\begin{array}{l}
\Big|V\left({\ttg}^{(n')}_j\om (j),{\ttg}^{(n')}_{j'}\om (j')\right)\\
\qquad +V\left({{\ttg}^{(n')}_j}^{-1}\om (j),
{{\ttg}^{(n')}_{j'}}^{-1}\om (j')\right)
-2V(\om (j),\om (j'))\Big|\\ \;\\
\qquad\qquad\leq  C\,
|\utheta|^2\left|\gamma (n',j)-\gamma (n',j')\right|^2
{\ov V}\,.\end{array}\eqno (4.2.10)$$
Here $C\in (0,\infty )$ is a constant, the upper bound ${\ov V}$
is taken from (2.3.2), and we use the notation from (4.2.7).

The bound (4.2.10) is crucial: this where
the structure of the group action is exploited.
It is based on the fact that the first-order terms
in the expansion in the LHS of (4.2.10) cancel each other,
due to the presence of elements ${\ttg}^{(n')}_j$ and
${\ttg}^{(n')}_{j'}$ and their inverses, ${{\ttg}^{(n')}_j}^{-1}$ and
${{\ttg}^{(n')}_{j'}}^{-1}$. This idea can be traced back to Pfister (1981)
and Fr\"ohlich and Pfister. (1981).

Further, the term
$|\gamma (n',j)-\gamma (n',j')|^2$ can be specified as
$$|\gamma (n',k)-\gamma (n',k')|^2=\begin{cases}0,
\;\hbox{ if }\;k,
k'\leq \ovr, \;\hbox{ or }\;\;k,k'\geq n',\\
\big[\vartheta (k-\ovr,n'-\ovr)\\
\quad -\vartheta (k'-\ovr,n'-\ovr)\big]^2,\\
\qquad\hbox{ if }\;\ovr<k,k'\leq n',\\
\vartheta (k-\ovr,n'-\ovr)^2,\\
\qquad\hbox{ if }\;\ovr<k\leq n',
k'\not\in ]\ovr,n'[,\\
\vartheta (k'-\ovr,n'-\ovr)^2,\\
\qquad\hbox{ if }\;\ovr<k'\leq n',
k\not\in ]\ovr,n'[ \end{cases}\eqno (4.2.11)$$
with notations $k={\td (j,s_0)}, k'={\td (j',s_0)}$.

The convexity property of the function exp, together with Eqn (4.2.10),
yield that, $\forall$ $a>1$,
$$\begin{array}{l}\diy\frac{a}{2}\exp\,\Big[-
h\Big({\bttg}_{n'}\big(
\obom (n)\vee
\bom (n'\setminus n)\big)|
\bOm^{\rm c} (n')\Big)\Big]\\
\quad +\diy\frac{a}{2}\exp\,\Big[-
h\Big({\bttg}_{n'}^{-1}\big(
\obom (n)\vee
\bom (n'\setminus n)\big)|
\bOm^{\rm c} (n')\Big)\Big]\\
\;\;\diy\geq a\exp\;\bigg[-\frac{1}{2}
h\Big({\bttg}_{n'}\big(
\obom (n)\vee
\bom (n'\setminus n)\big)|
\bOm^{\rm c} (n')\Big)\\
\qquad -\diy\frac{1}{2}
h\Big({\bttg}_{n'}^{-1}\big(
\obom (n)\vee
\bom (n'\setminus n)\big)|
\bOm^{\rm c} (n')\Big)\bigg]\\
\;\;\geq a\exp\,\Big[
-h\Big(
\obom (n)\vee\bom (n'\setminus n)|
\bOm^{\rm c} (n')\Big)\Big]
e^{-C\Phi /2}\,.\end{array}\eqno (4.2.12)$$
Here
$$\Phi =\Phi (n,{\ttg})=|\utheta|^2\sum_{(j,j')\in\cV_n\times\cV}
J({\td}(j,j'))\left|\gamma (n',k)-\gamma (n',k')
\right|^2.\eqno (4.2.13)$$
The series in (4.2.13) converges for $\sf P$ a.a. $T$,
owing to condition (2.3.5) and estimate (4.2.16) below.

The next observation is that
$$\begin{array}{l}\Phi\leq 3|\utheta |^2
\sum\limits_{(j,j')\in\cV_{n'}\times\cV}
{\mathbf 1}\big(k\leq k'
\big)J ({\td}(j,j'))\\ \;\\
\qquad\times\Big[\vartheta (k-\ovr,n'-\ovr) -
\vartheta (k'-\ovr,n'-\ovr)\Big]^2\end{array}\eqno (4.2.14)
$$
where, by virtue of the triangle inequality, for all
$j,j': k\leq k'$
$$\begin{array}{r}
0\leq\vartheta (k-\ovr,n'-\ovr) -
\vartheta (k'-\ovr,n'-\ovr)\qquad{}\\
\leq {\tt d}(j,j')\diy\frac{z(k-\ovr)}{Q(n'-\ovr)}.
\end{array}\eqno (4.2.15)$$
Thus,
$$\begin{array}{l}
\Phi\leq\diy\frac{3 |\utheta|^2}{Q(n'-\ovr)^2}\sum\limits_{(j,j')\in\cV_{n'}\times\cV}
J({\tt d}(j,j')){\tt d}(j,j')^2z(k-\ovr)^2\\
\quad\leq\diy\frac{3 |\utheta|^2}{Q(n'-\ovr)^2}
\Big[\sup_{j\in\cV}\sum\limits_{j'
\in\cV}J({\tt d}(j,j')){\tt d}(j,j')^2 \Big]\sum_{j\in\cV_{n'+\ovr}}
z(k-\ovr)^2.\end{array}
$$
Owing to (2.3.5), it remains to bound the sum $\sum_{j\in\cV_{n+\ovr}}
z(k-\ovr)^2$. Note that
$u(\ln u)^{1/2+\varepsilon}z(u)<1$ when $u\in (u_0(\varepsilon),\infty )$. Next, we use the
bound (4.1.1) on the number of vertices in $\cV_n\setminus\cV_{n-1}$.
Therefore,
$$\begin{array}{l}\sum\limits_{j\in\cV_{n'+\ovr}}
z(k-\ovr)^2=\sum\limits_{1\leq k\leq n'+\ovr}z(k-\ovr)
\sum\limits_{j\in\cV_k\setminus\cV_{k-1}}z(k-\ovr)\\
\qquad\qquad\leq C_0\sum\limits_{1\leq k\leq n'+\ovr}z(k-\ovr)
\leq C_1Q(n'-\ovr)\end{array}$$
and
$$\Phi\leq\frac{C(T)}{Q(n'-\ovr)}\to\infty,\;\hbox{ as }\;n'\to\infty.\eqno (4.2.16)$$
Hence, given $a>1$ for $n'$ large enough,
the term $ae^{-C\Phi /2}$ in the RHS of (4.2.12)
becomes $>1$. Consequently,
$$\begin{array}{l}\diy\frac{a}{2}\exp\,\Big[-
h\Big({\bttg}_{n'}\big(
\obom (n)\vee\bom(n'\setminus n)\big)
|\bOm^{\rm c} (n')\Big)\Big]\\
\quad +\diy\frac{a}{2}\exp\,\Big[-
h\Big({\bttg}_{n'}^{-1}\big(
\obom (n)\vee
\bom (n'\setminus n)\big)|
\bOm^{\rm c} (n')\Big)\Big]\\
\qquad\qquad\qquad
\;\;\geq\exp\,\Big[
-h\Big(\obom (n)\vee
\bom (n'\setminus n)|
\bOm^{\rm c} (n')\Big)\Big]\,.
\end{array}\eqno (4.2.17)$$
Eqn (4.2.17) implies that
$$\begin{array}{r}
q^{(n)}_{n'}
(\obom (n)|\bOm^{\rm c} (n'))=
\diy\int_{W^{\beta}_{\bx (n'\setminus n)}}{\rd}
\bom (n'\setminus n)
\quad\qquad{}\\
\diy\times\frac{\exp\big[-h
(\obom (n)\vee\bom (n'\setminus n)|\bOm^{\rm c} (n))
\big]}{\Xi_{\beta,n'}(
\bOm^{\rm c} (n'))},\end{array}
\eqno (4.2.18)$$
for any $n$ and $n'$ large enough obeys
$$\begin{array}{r}
a\Big[q^{(n)}_{n'}
({\ttg}\obom (n)|\bOm^{\rm c} (n'))
+q^{(n)}_{n'}
({\ttg}^{-1}\obom (n)|\bOm^{\rm c} (n'))\Big]\qquad{}\\
\geq 2q^{(n)}_{n'}
(\obom (n)|\bOm^{\rm c} (n'))\end{array}\eqno (4.2.19)$$
uniformly in the boundary condition $\bOm^{\rm c}(n')$.
Thus, (4.2.3) is established, which completes the proof of Theorem
\ref{Main2}. $\quad\Box$
\vskip 1 truecm

\subsection*{Acknowledgments}
This work has been conducted under Grant 2011/20133-0 provided by
the FAPESP, Grant 2011.5.764.45.0 provided by The Reitoria of the
Universidade de S\~{a}o Paulo and Grant 2012/04372-7
provided by the FAPESP. The authors
express their gratitude to NUMEC and IME, Universidade de S\~{a}o Paulo,
Brazil, for the warm hospitality.

\end{document}